\documentstyle[aps,prd,tighten,epsf]{revtex}
\begin{document}
\draft
\newcommand{\be}{\begin{equation}}
\newcommand{\ee}{\end{equation}}
\newcommand{\ben}{\begin{eqnarray}}
\newcommand{\een}{\end{eqnarray}}

\newcommand{\la}{{\lambda}}
\newcommand{\Om}{{\Omega}}
\newcommand{\ta}{{\tilde a}}
\newcommand{\bg}{{\bar g}}
\newcommand{\bh}{{\bar h}}
\newcommand{\si}{{\sigma}}
\newcommand{\th}{{\theta}}
\newcommand{\C}{{\cal C}}
\newcommand{\D}{{\cal D}}
\newcommand{\cA}{{\cal A}}
\newcommand{\cT}{{\cal T}}
\newcommand{\cO}{{\cal O}}
\newcommand{\eeo}{\cO ({1 \over E})}
\newcommand{\G}{{\cal G}}
\newcommand{\cL}{{\cal L}}
\newcommand{\T}{{\cal T}}
\newcommand{\M}{{\cal M}}

\newcommand{\p}{\partial}
\newcommand{\na}{\nabla}
\newcommand{\ssum}{\sum\limits_{i = 1}^3}
\newcommand{\dssum}{\sum\limits_{i = 1}^2}
\newcommand{\tal}{{\tilde \alpha}}

\newcommand{\tp}{{\tilde \phi}}
\newcommand{\tPhi}{\tilde \Phi}
\newcommand{\tpsi}{\tilde \psi}
\newcommand{\tim}{{\tilde \mu}}
\newcommand{\tr}{{\tilde \rho}}
\newcommand{\tir}{{\tilde r}}
\newcommand{\rp}{r_{+}}
\newcommand{\hr}{{\hat r}}
\newcommand{\rv}{{r_{v}}}
\newcommand{\dr}{{d \over d \hr}}
\newcommand{\dR}{{d \over d R}}

\newcommand{\hhf}{{\hat \phi}}
\newcommand{\hhM}{{\hat M}}
\newcommand{\hhQ}{{\hat Q}}
\newcommand{\hht}{{\hat t}}
\newcommand{\hhr}{{\hat r}}
\newcommand{\hhS}{{\hat \Sigma}}
\newcommand{\hhD}{{\hat \Delta}}
\newcommand{\hhm}{{\hat \mu}}
\newcommand{\hro}{{\hat \rho}}
\newcommand{\hhz}{{\hat z}}

\newcommand{\tD}{{\tilde D}}
\newcommand{\tB}{{\tilde B}}
\newcommand{\tV}{{\tilde V}}
\newcommand{\hT}{\hat T}
\newcommand{\tF}{\tilde F}
\newcommand{\tT}{\tilde T}
\newcommand{\hC}{\hat C}
\newcommand{\ep}{\epsilon}
\newcommand{\bep}{\bar \epsilon}
\newcommand{\ppp}{\varphi}
\newcommand{\Ga}{\Gamma}
\newcommand{\ga}{\gamma}
\newcommand{\hth}{\hat \theta}
\title{Dilaton Black Holes on Thick Branes}

\author{Marek Rogatko}
\address{Technical University of Lublin \protect \\
20-618 Lublin, Nadbystrzycka 40, Poland \protect \\
rogat@tytan.umcs.lublin.pl \protect \\
rogat@akropolis.pol.lublin.pl}
\date{\today}
\maketitle
\smallskip
\bigskip
\begin{abstract}
We study analytically a black hole domain wall system in dilatonic gravity
being the low-energy limit of the superstring theory. 
Using the C-metric construction
we derive the metric
for an infinitesimally thin domain wall intersecting dilaton black hole.
The behavior of the domain wall in the spacetime
of dilaton black hole was analyzed and it was revealed that the extreme
dilaton black hole always expelled the domain wall. 
We elaborated the
back reaction problem and concluded that topological kink solution smoothes out
singularity of the considered topological defect.
Finally we gave some comments concerning nucleation of dilaton black holes
on a domain wall and compared this process with the creation of static 
dilaton black holes in the presence of the domain wall.
We found that domain walls would rather prefer to nucleate 
small black holes on them, than large ones inside them.
\end{abstract}
\baselineskip=18pt
\par
\section{Introduction}
Motivated by the earlier investigations on uniqueness theorems for black
holes (see Ref.\cite{bl} and references therein) Wheeler coined the 
metaphoric dictum {\it black holes have no hair}. Regardless  of the
specific details of the collapse or the structure and properties of
the collapsing body a stationary black hole
emerged in the resultant process
which geometry was characterized by mass, charge and 
angular momentum.
Nowadays there has been a considerable resurgence of mathematical
works on black hole equilibrium states. The {\it no-hair} 
conjecture has been extended to the problem of nontrivial topology of
some fields configurations.
\par
The considerations presented in \cite{tl}
announced the existence of the Euclidean Einstein
equations corresponding to a vortex sitting on the horizon of the
black hole. In Refs.\cite{gg,cha,bon} numerical and analytical 
evidences for an Abelian-Higgs vortex acting as a long hair for 
Schwarzschild or Reissner-Nordstr\"om (RN) solution were established.
\par
The superstring theories are the most promising candidates for a 
consistent quantum theory of gravity.
Numerical studies of the low-energy string solutions revealed
that Einstein-dilaton black holes in the presence of a
Gauss-Bonnet type term were endowed with a nontrivial dilaton 
hair \cite{ka}. The extended moduli and dilaton hair connected with
axions for Kerr-Newmann black hole background were studied
in \cite{ka1}.
The dilaton black holes pierced by a thin vortex were intensively
studied both numerically and analytically \cite{mr,mr1,mr2,san}. It was shown
that the horizon of a charged dilaton black hole could support the long-range
fields of the Nielsen-Olesen vortex which could be considered as black hole hair.
It has been argued that the Euclidean dilaton black hole can support a
vortex solution sitting at the horizon \cite{mr}. Allowing the dilaton black hole to
approach extremality it was shown that the vortex was always expelled
from the extreme dilaton black hole. In the {\it thin string} limit
the metric of a conical dilaton black hole was obtained \cite{mr1,mr2,san}.\\
On their own
topological defects arising during spontaneous symmetry breaking
during phase transitions and their cosmological evolution play the
very important role in our understanding of the cosmological evolution
\cite{vil}.
Topological defects produced in the early stages of our universe
could shed some light on the high energy phenomena which were beyond
the range of our accelerators.\\
Recently domain walls were intensively studied due to the fact that our
universe might be a brane or defect being immersed in some higher dimensional
spacetime. The motivation for this fact comes from the unifications attempts
such superstring theories or M-theory \cite{bra}. This idea enables us to solve
the very intriguing phenomenological possibility of a resolution 
of the hierarchy problem.
\par
In this paper we shall try to provide some continuity
with our previous works \cite{mr,mr1,mr2}
and consider the problem of another topological defect, 
a domain wall and dilaton black hole.
We would like to find an
equivalent of black hole string solution given by Aryal {\it et al.} \cite{vi}
namely a domain wall dilaton black hole metric.\\
The paper is organized as follows. In Sec.II we 
derive an infinitesimal domain wall black hole metric
in dilaton gravity. In Sec.III
we analyze the fields equations of domain wall in the background of dilaton
black hole and dilaton C-metric. We derived an analytic {\it thin wall}
approximation useful in the back reaction problem. In Sec.IV we 
consider the problem of the expulsion of the domain wall by extremal
dilaton black hole. We gave analytical arguments that expulsion always
holds for this kind of black hole being the analog of the {\it Meissner effect}.
The same situation takes place for Abelian-Higgs vortex and extremal dilaton
black hole. In Sec.V we deal with the gravitational back reaction and conclude
that topological kink solution smoothes out the singularity of the domain wall.
In Sec.VI we study the problem of a nucleation of dilaton black holes on
a domain wall and the process of nucleation of static dilaton black hole
pairs in the presence of a domain wall. Sec.VII concludes our results.

\section{The Dilaton Black hole-wall metric}
In this section we will try to find an equivalent of a black
hole string solution given by Aryal {\it et al} in \cite{vi}, i.e.,
an infinitesimally thin domain wall
with a dilaton black hole being the static spherically symmetric
solution in dilaton gravity. Dilaton gravity is the low-energy
limit of the superstring theory. The action of this theory is given by \cite{gro}
\be
S = \int dx^4 {\sqrt{-g} \over 16 \pi} \bigg[ R - 2 \big( \na \phi \big)^2
- e^{-2 a \phi}F^2 \bigg].
\ee
The equations of motion derived from the variational principle may be 
written as follows:
\ben
\na_{\mu} \bigg(  e^{-2 a \phi}F_{\mu \nu} \bigg) = 0, \\ \label{q}  
\na_{\mu}\na^{\mu} \phi + {a \over 2} e^{-2 a \phi}F^2 = 0, \\ \label{f}
G_{\mu \nu} = T_{\mu \nu}(\phi, F),
\een
where the energy-momentum tensor yields
\be
 T_{\mu \nu}(\phi, F) =  e^{-2 a \phi} \big( 4 F_{\mu \rho}
F_{\nu}{}{}^{\rho} - g_{\mu \nu}F^2 \big)
- 2g_{\mu \nu}(\na \phi)^2 + 4 \na_{\mu} \phi \na_{\nu} \phi.
\ee
In the case of RN black hole-domain wall
system the metric was derived 
in Ref.\cite{egs}. It constituted the extension of the results 
obtained in \cite{emp1,emp2}. The key point in the above derivation will be the notion 
of the C-metric being an axially symmetric solution of Einstein gravity 
which represents two black holes uniformly accelerating apart. The force
for acceleration is provided by a conical excess between black holes or by a 
conical deficit (string) extending from each black hole to infinity. An external
gravitational field can remove the nodal singularities \cite{ern1}. If the black holes
are electrically (magnetically) charged the cause of the acceleration is
electric (magnetic) \cite{ern2}.\\
We begin with the generalization of the C-metric in dilaton gravity given by \cite{cc}:
\be
ds^2 = {1 \over A^2 (x - y)^2} \bigg[
F(x) \bigg( G(y)dt^2 - {dy^2 \over G(y)} \bigg) +
F(y) \bigg( {dx^2 \over G(x) } + G(x) d\phi^2 \bigg)
\bigg],
\label{cc}
\ee
where we have denoted
\ben
e^{-2a \phi} &=& {F(y) \over F(x)}, \qquad 
F(\xi) = \big( 1 + r_{-}A \xi \big)^{2a^2 \over 1 + a^2}, \qquad
A_{\phi} = q x, \\ \nonumber
G(\xi) &=& \big[ 1 - \xi^2 \big( 1 + \rp A \xi \big) \big] 
\big( 1+ r_{-} A \xi \big)^{1 - a^2 \over 1 + a^2}.
\een
The metric (\ref{cc}) has two Killing vectors ${\p \over \p t}$ and
${\p \over \p \phi}$.
The norm of the Killing vector ${\p \over \p \phi}$ vanishes
at $x = \xi_{3}$ and $x = \xi_{4}$, this fact corresponds to the existence
of the poles of the spheres surrounding the black holes. The axis $x = \xi_{3}$
points along the symmetry axis towards spatial infinity, while the axis
$x = \xi_{4}$ points towards the other black hole.\\
For $\rp A < 2/3\sqrt{3}$ the function
$G(\xi)$ has four real roots which
in ascending order are denoted by $\xi_{2}~, \xi_{3}~, \xi_{4}$
and we define $\xi_{1} = - {1 \over r_{-} A}$. The surface $y = \xi_{1}$
is singular for $a > 0$. It is analogous to the singular
surface (the inner horizon of the dilaton black hole). The surface $y = \xi_{2}$
is the black hole horizon, while $y = \xi_{3}$ is the acceleration horizon
for an observer comoving with the black hole.
These two surfaces are both Killing horizons for ${\p \over \p t}$.
In the limit $\rp A \ll 1$ and $r_{-} A \ll 1$ one obtains
\ben
\xi_{1} = - {1 \over \rp A} + O(A), \qquad
\xi_{2} = - {1 \over \rp A} + O(A), \\ \nonumber
\xi_{3} = - 1 - {\rp A \over 2} + O(A^2), \qquad
\xi_{4} = 1 - {\rp A \over 2} + O(A^2).
\een
As was discussed in Ref.\cite{cc} in the ordinary C-metric is impossible
to choose generally such a range of $\phi$ that the metric (\ref{cc})
is regular at both $x = \xi_{3}$ and $x = \xi_{4}$. One can get rid of
the nodal singularity at $x = \xi_{4}$ by choosing $\phi \in [0, {4\pi \over
\mid G'(\xi_{4}) \mid} ]$, but then there is a positive deficit angle
running the $\xi_{3}$ direction. It has the interpretation as a string 
with a positive mass per unit length $\mu = 1 - {\mid G'(\xi_{3}) \mid \over
\mid G'(\xi_{4}) \mid}$
pulling
the accelerating dilaton black holes away to infinity. 
On the other hand choosing $\phi \in [0, {4\pi \over
\mid G'(\xi_{3}) \mid} ]$, means that one has a negative deficit
angle along $\xi_{4}$ direction, interpreted as the black holes are
being pushed apart by a {\it rod} of the negative mass per unit length
equals to $\mu = 1 - {\mid G'(\xi_{4}) \mid \over
\mid G'(\xi_{3}) \mid}$.
\par
In order to construct the wall-dilaton black hole metric we shall follow the Israel
procedure \cite{isr}, i.e., the discontinuity of the extrinsic curvature is provided by the
tension $\sigma$ of a domain wall. Thus, it implies the following:
\be
\big[ K_{ij} \big] = 4 \pi G \sigma h_{ij},
\ee
where $h_{ij}$ is the metric induced on the wall. 
Having in mind \cite{egs}, an
appropriate umbilic surface can be found at $x = 0$. This has
normal $n = {1 \over Ay} dx$, and the induced metric is of the form
\be
ds^2 = {1 \over A^2 y^2} \bigg[ G(y) dt^2 - {dy^2 \over G(y)} + 
F(y)d\phi^2 \bigg].
\ee
The extrinsic curvature in this case is $K_{ij} = A h_{ij}$ and the
Israel condition implies that the domain wall tension is equal to $\sigma = 
A/ 2 \pi G$.
\par
One can choose the conical singularity to lie at $x = \xi_{3}$, on the side
$x < 0$ of this surface. However the string will vanish from the spacetime 
if we take two copies of the side $x > 0$ and glue them together along 
$x = 0$. This construction is equivalent to determining $\mid x \mid$
for $x $ in the metric (\ref{cc}).  
The metric induced on the domain wall has an interesting form after introducing
new radial and time coordinates
of the form $r = - {1 \over A y}$ and  $T = {t \over A}$, namely
it reduces to
\be
ds^2 = - \bigg(
1 - {\rp \over r} - A^2 r^2 \bigg) 
\bigg( 1 - {r_{-} \over r} \bigg)^{1 - a^2 \over 1 + a^2}dT^2 +
{dr^2 \over 
\bigg(
1 - {\rp \over r} - A^2 r^2 \bigg) 
\bigg( 1 - {r_{-} \over r} \bigg)^{1 - a^2 \over 1 + a^2}} +
r^2 \bigg( 1 - {r_{-} \over r} \bigg)^{2 a^2 \over 1 + a^2} d\phi^2.
\ee
As in Ref.\cite{egs} we have chosen the conical singularity to be at $x = \xi_{3}$
on the side $x < 0$. The consequence of this is that if one takes 
two copies of the side $x > 0$ and glue them along $x = 0$, the string
disappears from the spacetime. \\
The charge of the black hole can be measured by integrating the flux on
a sphere surrounding it. It implies
\ben
Q &=& 2 {1 \over 4 \pi} \int dx d\phi F_{x \phi} = {\Delta \phi \over 2 \pi}
\big( A_{\phi}(x = \xi_{4}) -  A_{\phi}(x = 0) \big) \\ \nonumber
&=& {2 \xi_{4} \over
q A^2 \big( r_{-} A \big)^{- 2a^2 \over 1 + a^2} (\xi_{4} - \xi_{3})
(\xi_{4} - \xi_{2})(\xi_{4} - \xi_{1})^{1 - a^2 \over 1 + a^2} (1 + a^2) },
\een
where $\Delta \phi = {4\pi \over
\mid G'(\xi_{4}) \mid}$ is the period of $\phi$ coordinate.\\
As was pointed out in \cite{egs} the constructed domain wall contained two
black holes at antipodal points of a spherical domain wall.
The constructed black hole will neither swallow up the brane nor slide off
of it \cite{rad}. The letter fact was revealed because of acting the
elastic restoring force by means of which the brane acted on the 
dilaton black hole.

\section{Domain wall black holes }
In this section we shall describe 
behavior of the domain wall in the spacetime of dilaton black hole.
A static, spherically symmetric solution of the equations
of motion derived from the action $S$ is concerned
it is determined by the metric of a
charged dilaton black hole. The metric may be written as
\cite{db}
\be
ds^2 = - \left ( 1 - {2 M \over r} \right ) dt^2 +
{ d r^2 \over \left ( 1 - {2 M \over r} \right ) } + r \left ( 
r - {Q^2
\over M} \right ) (d \theta^2
+ \sin^2 \theta  d \ppp^2),
\ee
where we define $r_{+} = 2M$ and $r_{-} = {Q^2 \over M}$ which
are related to the mass $M$ and charge $Q$ by the relation $Q^2 =
{r_{+} r_{-} \over 2} e^{2 \phi_{0}}$. The charge of the dilaton
black hole $Q$, couples to the field $F_{\alpha \beta}$.
The dilaton field is given by $e^{2\phi} = \left ( 1 - {r_{-} \over r} \right )
e^{-2\phi_{0}}$, where $\phi_{0}$ is the
dilaton's value at $r \rightarrow \infty$. The event horizon is
located at $r = r_{+}$. For $r = r_{-}$ is another
singularity, one can however ignore it because $r_{-} < r_{+}$. The extremal black hole
occurs when $r_{-} = r_{+}$,
when $Q^2 = 2M^2 e^{2\phi_{0}}$.
\par
We consider a general matter Lagrangian with real Higgs field and the symmetry
breaking potential of the form as follows:
\be
{\cal L}_{dw} = - {1 \over 2} \na_{\mu} \varphi \na^{\mu} \varphi - U(\varphi).
\ee
The symmetry breaking potential $U(\varphi)$ has a discrete set of degenerate minima.
The energy-momentum tensor for the domain wall yields
\be
T_{ij}(\varphi) = - {1 \over 2} g_{ij} \na_{m} \varphi \na^{m} \varphi
- U(\varphi) g_{ij} + \na_{i} \varphi \na_{j} \varphi.
\ee
For the convenience we scale out parameters via transformation $ X = {\varphi / \eta}$
and $\ep = 8 \pi G \eta^2$. The parameter $\ep$ represents the gravitational 
strength and is connected with the gravitational interaction of the Higgs field.
Defining
$V(X) = {U(\varphi) \over V_{F}}$, where $V_{F} = \lambda \eta^4$ we arrive
at the following expression:
\be
8 \pi G {\cal L}_{dw} = - {\ep \over w^2} \bigg[
w^2 {\na_{\mu} X \na^{\mu} X \over 2} + V(X) \bigg],
\label{dw}
\ee
where
 $w = \sqrt{{\ep \over 8 \pi G V_{F}}}$
represents
the inverse mass of the scalar  after symmetry breaking, 
which also characterize the width of the wall
defect within the theory under consideration.
Having in mind (\ref{dw}) the equations for $X$ field may be written as follows:
\be
\na_{\mu} \na^{\mu} X - { \p V \over \p X} = 0,
\ee
where without loss of generality we have set $w = 1$ in order to fix our unit.
In the background of the dilaton black hole spacetime
the equation of motion for the scalar field $X$ yields
\be
{1 \over r \left ( r - {Q^2 \over M} \right )}
\p_{r} \bigg[ \big ( r - {Q^2 \over M} \big) \big( r - 2 M \big)
\p_{r} X \bigg] + 
{1 \over r \left ( r - {Q^2 \over M} \right ) \sin \theta }
\p_{\theta} \bigg[ \sin \theta \p_{\theta} X \bigg] =
{ \p V \over \p X}.
\label{xxxx}
\ee
As in the case of the vortex and black hole \cite{tl,gg,mr1,mr2} the fields were 
approximated as functions of $\sqrt{g_{33}}$. Now, we guess the ansatz 
$X(z) = X(r \cos \theta)$. Then, we can establish the following:
\be
\na_{\mu} \na^{\mu} X = X'' - {2 M z^2 \over r^3} X'' +
{X' \over z} - {2 M \over r z} X'.
\ee
Taking into account the fact that outside the black hole
horizon $r$ is far more greater than $M$, and assuming
that the thickness of the wall is much less than the black hole
horizon, i.e., $M \gg 1$ one can deduce that $X$ is approaching
the flat space solution. Thus in the {\it thin wall} approximation
the thin wall can be painted on a dilaton black hole. In the case of
Schwarzshild solution this fact was confirmed by the numerical calculations
\cite{jap}.
Preliminary numerical studies in the dilaton black hole 
case also confirmed this analytic results \cite{prep}.
\par
Now we proceed to the problem of painting the domain wall onto the dilaton
C-metric.
As we have expected our gravitating wall-black hole system will
be described by the dilaton C-metric. By virtue of the new variables 
defined by
\be
r = - {1 \over A y}, \qquad
T = {t \over A}, \qquad
\theta = \int_{x}^{x_{3}}{dx \over \sqrt{G(x)}},
\ee
one can reach to the following metric:
\be
ds^2 = {1 \over \big ( 1 + A r x \big )^2}
\bigg[ - F(x) H(r) dT^2 + {F(x) \over H(r)} dr^2 +
K^2(r) \big( d \theta^2 + G(x) d\phi^2 \big) \bigg],
\ee
where we have denoted
\ben
H(r) = - \bigg( 1 - {\rp  \over r} - A^2 r^2 \bigg)
\bigg( 1 - {\rp \over r} \bigg)^{1- a^2 \over 1 + a^2},\\ \nonumber
K^2(r) = r^2 \bigg( 1 - {r_{-} \over r} \bigg)^{2 a^2 \over 1 + a^2}.
\een
Now it is easily seen that the variable $x$ is 
basically $\cos \theta$ and as in Ref.\cite{egs} we guess $z = {x \over A y}$.
Thus, after straightforward but tedious calculations we have:
\ben
\na_{\mu} \na^{\mu} X &=&
X'' \big( z A - 1 \big)^2 \bigg[
{G(x) \over F(y)} - {A^2 z^2 G(y) \over F(x)} \bigg] \\ \nonumber
&+&
X' A \big( z A - 1 \big) \bigg[
(z A - 1 ) \bigg ( \big( F(x) G(x) \big)' y + \big( F(y) G(y) \big)' x 
\bigg) - 2 \bigg( {G(x) \over F(y)} + {A G(y) \over F(x)} \bigg) -
{2 z A ( z A - 1 ) G(y) \over F(x)}
\bigg].
\een
The {\it thin wall} approximation \cite{egs} in the context of the C-metric means that
$A \mid \xi_{2} \mid \ll 1$, i.e., the black hole horizon radius has to be large.
However for a self-gravitating domain wall there is a limit due to the wall formation.
This limit is given by the size of the spontaneously compactified
spacetime which corresponds to the acceleration horizon, then we will
work having in mind the large regime of the accelerated horizon, i.e.,
$A \mid \xi_{3} \mid \ll 1$. 
In this case the wall fields differ significantly from their vacuum because
$z \sim 1$. On the other hand, the values of $y$ are bounded by the black hole
and the acceleration horizon $1 < \mid \xi_{3} \mid \le y < \mid \xi_{2} \mid $
and therefore $x \le A y \le 1$.
Thus we have
\be
\Box X = X'' + O (A).
\label{xx}
\ee
We complete this section by the conclusion
that the flat spacetime solution $X_{0}(z)$ is
a good approximation to the solutions of the field equations in the 
dilaton C-metric spacetime.

\section{Expelling of the wall by the extremal dilaton black hole}
In the previous section we argued the existence of the domain wall solution
in the case of a large black hole masses.
Now we shall consider the case of a very small extremal black hole
sitting inside the domain wall. Inside the core of the wall
the potential term is very small compared to the gradient terms, so we
can neglect it. The solution for a wall in the absence of black hole in the region 
adjacent to its core is
$X \simeq z - z_{0}$, we try the ansatz $X(r, \theta) = b(r) \cos \theta$
\cite{egs}. Applying the ansatz to Eq.(\ref{xxxx})
one gets
\be
2 \big( r - M - {Q^2 \over 2 M} \big) b' + \big( r - {Q^2 \over M} \big)
\big( r - 2 M \big) b'' - 2 b = 0.
\label{bla}
\ee
The solution of Eq. (\ref{bla}) is provided by
\be
X \approx \big( r - M - {Q^2 \over 2 M} \big) \cos \theta.
\ee
One can observe that if we have to do with a non-extreme dilaton black hole, one
has $X \ne 0$ on the horizon.
For an extermal dilaton black hole for which
$\rp = r_{-}$ occurs, i.e.,
$ Q = \sqrt{2} M$ we obtain that
$b(\rp) = 0$ and $X(\rp) = 0$.
This is in accord with the fact of the expulsion of a
domain
wall by a small extremal dilaton black hole sitting inside the domain wall.
The same situation was revealed in the case of the extreme dilaton black hole and
the Nielsen-Olesen vortex \cite{mr1,mr2}. 
We gave analytical and numerical arguments that the vortex was always expelled
from the considered black hole. In the case of an Abelian-Higgs vortex
and extremal dilaton black hole system the analog of the {\it Meissner effect}
was found.
\par
After proving that the flux expulsion must take place for a sufficiently thick
domain wall from the extremal dilaton black hole, we treat the case of a thin
domain wall. As was shown in Ref.\cite{ext} the metric of the extreme dilaton black hole 
near horizon
may be written in the form of Bertotti-Robinson metric, namely
\be
ds^2 = -{ \rho^2 \over M^2 - \Sigma^2}dt^2 + {M^2 - \Sigma^2 \over \rho^2} d\rho^2 +
(M^2 - \Sigma^2) \big( d \theta^2 + \sin^2 \theta d \phi^2 \big),
\ee 
where $\rho^2 = x_{i}x^{i}, x_{i}$ are isotropic coordinates and
$\Sigma^2 = {1 \over 2}Q^2$. The equation of motion for
 $X$ field implies
\be
\p_{\rho} \big( \rho^2 \p_{\rho} X \big) + X_{,\theta \theta} + 
\cot \theta X_{, \theta} = X \big( X^2 - 1 \big) \big(  M^2 - \Sigma^2 \big).
\label{ext1}
\ee
Let us assume that there is a flux expulsion. Then on the horizon, where 
$\rho \rightarrow 0$, one has $X = 0$ and  $(M^2 - \Sigma^2) X^3 \ll 1$.
Having all these in mind and we integrate (\ref{ext1}) on the interval
$(\theta, {\pi \over 2})$, for $\theta > \beta_{0}$. We arrive at
the inequality of the form
\be
\p_{\theta} X(\theta) > \cot \theta X (M^2 - \Sigma^2).
\ee
Then using the fact that $X_{,\theta \theta} < 0$ on $[0, {\pi \over 2}]$
we deduce that 
\be
X_{,\theta \theta} < {X(\theta) - X(\theta_{0})  \over \theta - \theta_{0}} <
{X(\theta) \over \theta - \theta_{0}} < {X(\theta) \over \theta - \beta}.
\ee
This enables us to write the following:
\be
{1 \over M^2 - \Sigma^2} > \big( \theta - \beta \big) \cot \theta.
\label{in}
\ee
The above relation
must hold over the range of $\theta \in (\beta, {\pi \over 2})$ for
the expulsion to occur. Since $\theta - \beta > 0$, $\cot \theta$ on this
interval is greater than zero, then the relation (\ref{in}) always holds and
one gets the expulsion of the thin domain wall from the extremal dilaton
black hole.
\section{Gravitational Back reaction}
In order to study the gravitational back reaction problem, we shall
consider the thick-wall dilaton black hole metric (\ref{cc}). We denote for simplicity
$\Omega = A (x - y)$ and perform a linearized calculations in $\ep = 3A/2$ as
in Ref.\cite{egs}, writing $\Omega = \Omega_{0} + A \Omega_{1}$ and so on. Near the core
of the domain wall $\Omega_{1}/\Omega_{0} = O(1)$ and tends to zero far away
from it.
Let us calculate
\ben
g^{xx}X_{,x}X_{,x} = {A^{2} (x - y)^2 \bigg[ 1 - x^2 \big( 1 + \rp A x \big) \bigg]
\big( 1 + r_{-} A x \big)^{1 - a^2 \over 1 + a^2}  
\over \big( 1 + r_{-} A y \big)^{2 a^2 \over 1 + a^2}} {1 \over A^2 y^2} (X')^2 
\simeq (X_{0}')^2, \\
g^{yy}X_{,y}X_{,y} = { - A^{2} (x - y)^2 \bigg[ 1 - y^2 \big( 1 + \rp A y \big) \bigg]
\big( 1 + r_{-} A y \big)^{1 - a^2 \over 1 + a^2}  
\over \big( 1 + r_{-} A x\big)^{2 a^2 \over 1 + a^2}} \big(
- {2 \over y} X' \big)^2 
\simeq  O(A^2),
\een 
and 
\ben
g^{xx}\phi_{,x}\phi_{,x} = {A^{2} (x - y)^2 \bigg[ 1 - x^2 \big( 1 + \rp A x \big) \bigg]
\big( 1 + r_{-} A x \big)^{1 - a^2 \over 1 + a^2}  
\over \big( 1 + r_{-} A y \big)^{2 a^2 \over 1 + a^2}} {1 \over A^2 y^2} (\phi')^2 
\simeq (\phi_{0}')^2, \\
g^{yy}\phi_{,y}\phi_{,y} = { - A^{2} (x - y)^2 \bigg[ 1 - y^2 \big( 1 + \rp A y \big) \bigg]
\big( 1 + r_{-} A y \big)^{1 - a^2 \over 1 + a^2}  
\over \big( 1 + r_{-} A x\big)^{2 a^2 \over 1 + a^2}} \big(
- {2 \over y} \phi' \big)^2 
\simeq  O(A^2).
\een 

From now on, for simplicity, we set $a = 1$ in our considerations. Then,
equations of motion (\ref{q}) and (\ref{f}) take the forms as follows:
\be
\p_{x} \bigg[ e^{- 2 \phi} {F(x) F(y) \over \Omega^4} F^{\phi x} \bigg]
+
\p_{y} \bigg[ e^{- 2 \phi} {F(x) F(y) \over \Omega^4} F^{\phi y} \bigg] =
0,
\label{qq}
\ee
\be
\p_{x} \bigg[ {F(x) G(x) \over \Omega^4} \phi_{, x} \bigg] -
\p_{y} \bigg[ {F(y) G(y) \over \Omega^4} \phi_{, y} \bigg] +
{1 \over 2}\sqrt{- g} e^{-2 \phi} F^2 = 0,
\label{ff}
\ee
As in Ref.\cite{mr2} we shall assume that the first order perturbed solutions 
are determined by
\be
\phi_{1} = \phi_{1}(z), \qquad A_{\mu}^{(1)} = g(z) A_{\mu}^{(0)}.
\ee
Taking into account (\ref{xx}) and (\ref{ff}) we draw a conclusion
that to the leading order in $A$ we get $\phi_{1} = const$. Next from
the relation (\ref{qq}) we have
\be
- 2 \p_{x} \big( \phi_{1} q \big) + \p_{x}^2 
\bigg[ g(z) A_{\phi}^{(0)} \bigg] = 0.
\ee
Then, one gets that $g(z)  = const$. The gauge potential for the Maxwell  fields
is unaltered by the presence of the domain wall.\\
To the leading order in $A$ we have the following generalized Einstein 
equations:
\ben
R_{0}{}{}^{0} &=& - 2 e^{- 2 \phi} q^2 \Omega^4 + \ep V(X),\\
R_{x}{}{}^{x} &=& 3 e^{-2 \phi} q^2 \Omega^4  + 4 (\phi')^2 + \ep V(X) + 
\ep (X_{0}')^2, \\
R_{y}{}{}^{y} &=&  - 2 e^{-2 \phi}q^2 \Omega^4 + \ep V(X) + O(A), \\
R_{\phi}{}{}^{\phi} &=&  6 e^{-2 \phi}q^2 \Omega^4 + \ep V(X), \\
R_{xy} &=& - {4 z \over A y^2} \big( \phi' \big)^2
- {2 \ep z \over A y^2} \big( X_{0}' \big)^2.
\een
As was mentioned in \cite{egs} because of the fact that the variation
of the extrinsic curvature due to the wall is carried
by $\Omega$, one guesses that $F$ and $G$ will effectively 
take their background values. After lengthy calculations 
we find the following:
\ben
R_{0}{}{}^{0} - R_{y}{}{}^{y} &=&
{1 \over 2 F(x) F(y)^2} \bigg[ G(y) \Omega^2 F'(y)^2 + 4 G(y) 
\Omega \Omega_{,yy} F(y)^2 \bigg], \\
R_{\phi}{}{}^{\phi} - R_{x}{}{}^{x} &=&
{1 \over 2 F(x)^2 F(y)} \bigg[
- G(x) \Omega F'(x)^2 - 4 G(x) F(x)^2 \Omega \Omega_{,xx} \bigg], \\
R_{0}{}{}^{0} - R_{\phi}{}{}^{\phi} &=&
{1 \over 2 F(x) F(y)} \bigg[ F(x) \Omega^2 G''(x) +
F(y) \Omega^2 G''(y) + 2 G(y) F'(y) \Omega \Omega_{,y} \\ \nonumber
&-& 2 F(y) G'(y) \Omega \Omega_{,y} + 2 G(x) F'(x) \Omega \Omega_{,x}
- 2 F(x) G'(x) \Omega \Omega_{,x} \bigg].
\een
The above relations suggest that $\Omega$ may be written as
\be
\Omega = A \big( f - y \big),
\ee
where $ f_{0} = \mid x \mid$. Inputting the ansatz for
$\Omega$ and $X_{0} = \tanh z$ one obtains
\be
R_{0}{}{}^{0} - R_{y}{}{}^{y} =
{1 \over 2 F(x) F^2(y)} \bigg[ G(y) A^2 (f - y)^2 (r_{-}A)^2 + 
4 G(y) A^2 y \big( -1 + A \mid z \mid \big) f_{,yy} F^2(y) \bigg],
\ee
\be
R_{\phi}{}{}^{\phi} - R_{x}{}{}^{x} =
{1 \over 2 F^2(x) F(y)} \bigg[ - 
G(x) A^2 (f - y)^2 + 4 G(x) F^2(x) A^2 y
\big( 1 - A \mid z \mid \big) f_{,xx} \bigg] = 
- 4 \phi'^2 
- {\ep \over \cosh ({x \over Ay})},
\ee
\ben
R_{0}{}{}^{0} - R_{\phi}{}{}^{\phi} &=&
{1 \over 2 F(x) F(y)} \bigg[ - 2 F(x) A^2 (f - y)^2 \big(1 + 3 \rp A x \big)
- 2 F(y) A^2 (f - y)^2 \big(1 + 3 \rp A y \big) \\ \nonumber
&+& 2 G(y) r_{-} A^3 y \big( A \mid z \mid - 1 \big) (f_{,y} - 1)
+ 2 F(y) A^2 y^2 \big(2 + 3 \rp A y \big) \big(A \mid z \mid - 1 \big)
(f_{,y} - 1) \\ \nonumber
&+& 2 G(x) r_{-}A^2 y f_{,x} \big( A \mid z \mid - 1 \big) +
2 F(x) A^2 y x \big( 2 + 3 \rp A x \big) f_{,x} \big( A \mid z \mid - 1 \big)
\bigg].
\een
Since $f_{,x} = O(A)$ and $f_{,y} = O(A^2)$ the generalized Einstein's
equations in dilaton gravity are satisfied to the leading order in $A$.
\par
Thus, we get the solution
\be
f = - \ep {y \over 2} \bigg[  {1 \over 6} \tanh^2 \bigg( {x \over Ay} \bigg)
+ {2 \over 3} \ln \cosh \bigg( {x \over Ay} \bigg) \bigg]
- 2 A \int dx'dx {z \over x'} \bigg( {d \phi \over dx'} \bigg)^2.
\ee 
As in the case of the domain wall black hole system in general relativity
\cite{egs}
the topological kink solution smoothes out the shell-like
singularity of the infinitisemal domain wall. The same situation
takes place for topological vortex solutions which smooth 
out the delta function singularity in various kind of metrics \cite{tl,gg,mr,mr1,mr2}.
\section{Nucleation of dilaton black holes on and in the presence of domain walls}
\par
In this section we shall be concerned with the process
of nucleation of dilaton black holes on the domain wall and black holes
enclosed by a wall.
Let us first consider the case of dilaton black hole on a domain wall.
The probability of nucleation of a domain wall with a black hole on it
is given by $exp [ -(I - I_{0})]$, where $I_{0}$  is the Euclidean action
of the initial configuration, while $I$ is the Euclidean action of the final
state with dilaton black hole on it. We assume the no-boundary conditions
for the wave function of the considered universe. The considered exponent
can be also viewed as the ratio of probabilities to nucleate a domain wall
with or without black hole on it. In order to construct the wall black 
hole instanton one can use the on-shell equation \cite{haw}
\be
I = - {1 \over 4} \bigg( {\cal A}_{acc} +{\cal A}_{bh} \bigg),
\ee
giving the action in terms of the appropriate area of the horizons.\\
One obtains the following:
\be
I = - {\xi_{4} \over 2 \pi \sigma^2 \mid G'(\xi_{4}) \mid}
\bigg[ 
{1 \over (\xi_{3} - \xi_{4}) \xi_{3}} \bigg( 1 + r_{-}A \xi_{3} 
\bigg)^{2 a^2 \over 1 + a^2}
+ 
{1 \over (\xi_{2} - \xi_{4}) \xi_{2}} \bigg( 1 + r_{-}A \xi_{2} 
\bigg)^{2 a^2 \over 1 + a^2}
\bigg],
\label{soll}
\ee
while the Euclidean action for a domain wall is given by \cite{cal}
\be
I_{0} = - {1 \over 8 \pi \sigma^2}.
\ee
If one considers the case of $a = 1$ and the limit $\rp \ll 1$
and $r_{-}A \ll 1$ the expression (\ref{soll}) simplifies and the resulting 
relation yields
\be
I = - {1 \over 8 \pi \sigma^2}
\bigg[
1 - 2 \pi \sigma \bigg( 4 M + {Q^2 \over M} \bigg)
\bigg] + O(A^2).
\ee
Hence,
\be
I - I_{0} = \big( I - I_{0} \big)_{RN-dw} + {Q^2 \over 4 \sigma M},
\ee
where $\big( I - I_{0} \big)_{RN-dw}$ is the value of the exponent in the nucleation
rate for the RN-domain wall system. In the extremal dilaton black hole
case it yields
\be
(I - I_{0})_{ext} = 1.5 {M \over \sigma}.
\ee
The black hole mass $M$ is a parameter which can be varied independently of the domain
wall tension. Therefore it can be arbitrary small. 
\par
In addition to the process of nucleation of dilaton black holes on the wall
there exists the process of nucleation of black holes enclosed by a domain wall.
In Ref.\cite{cal}
it was found that domain walls could nucleate black holes at a finite distance
from them. The double-sided nature of the domain wall caused that it enclosed
a black hole on each side of it.
The authors, among all, considered both charged and
neutral black holes, while Ref.\cite{cal1} was devoted to the pair creation of 
black holes in the presence of supergravity domain walls with broken and unbroken
symmetry.
Of course, the tantalizing question arises, which of these two 
processes is more likely to take place.
\par
In order to build the instanton for static black hole nucleation one should first 
construct 
the Lorentzian section. To get a nonzero probability it ought to be required that a
spatially closed universe has a finite three-volume (it caused that the total energy
at the instant of nucleation vanished). This cut-and-paste procedure is depicted on
Fig. 2 in Ref.\cite{cal}. For the reader's convenience we quote this procedure, namely,
one has two copies of dilaton black hole spacetime and join them along $r = const$ timelike
hypersurface at the location of the domain wall. Then, one identifies two external 
regions of the dilaton spacetime (this implies the $R \times S^{1} \times S^{2}$ topology
of the spacetime). The resulting spacetime contains two domain walls and
two domains containing dilaton black hole in each. In order to get the domain wall
and dilaton black hole instanton
one starts with the usual Riemannian section of dilaton
black hole with mass $M$, with topology $S^{2} \times R^{2}$ which will be cut 
along $r = const$ and glue back to back. The outcome will be a {\it baguette}
with $S^{2} \times S^{2}$ topology with a {\it ridge} at the domain wall. As in the
Schwarzschild case one may check that if we take the hypersurface $r = 3 M$, it will be 
totally umbilic, its extrinsic curvature
$K_{ij}$ is proportional to induced metric $h_{ij}$ on the domain wall and 
$[K_{ij}] = 4 \pi \sigma  h_{ij}$. In the end the Riemannian 
space has been joined to the Lorentzian spacetime depicting the creation 
from nothing of a closed spacetime containing two domain walls and two 
domains containing a dilaton black hole in each of them.
\par
Now we shall proceed to the problem of creation of static 
magnetically charged dilaton 
black holes. 
The static black hole is the black hole which attractive gravitational energy
exactly counterbalances the repulsive energy of the wall. We pointed out
that the {\it static limit} domain wall are the wall for which $ {\dot{r}}~ = 0$,
where the derivative was taken with respect to
the proper time of the wall. The static 
domain wall
lies at $r_{st} = 3 M$ \cite{cal}.
The case of electrically charged dilaton black hole is quite analogous,
except the fact that the electromagnetic charge must be imaginary on the 
Riemannian section \cite{hawr}.
The Euclidean action for the instanton will include an electromagnetic and
dilaton contributions. Thus, we obtain
\be
I_{st} = I_{E_{1}} + I_{E_{2}} + I_{E_{3}} =
-{1 \over 2} \int_{W} d^3 x  \sqrt{h} +
\int_{M} d^4 x
{\sqrt{g} \over 16 \pi} \bigg[
2 (\na \phi)^2 + e^{- 2 \phi} F^2 \bigg].
\label{eu}
\ee
There are no boundary terms because the considered instantons are
compact and without boundary.
Calculating the first term in (\ref{eu}), due to the presence of the domain wall
\cite{cal}, gives
\be
I_{E_{1}} = 4 \pi \sqrt{ f(r)} \beta R(r)^2 \mid_{r_{st}},
\ee
where $R(r) = \sqrt{r \left ( 
r - {Q^2
\over M} \right )}$ and 
$f(r)$ is the $g_{00}$ Euclidean dilaton black hole coefficient
in $(t, r)$ coordinates.
The action is evaluated at $r_{st}$ and $ \beta = 8 \pi M$ is the
instanton period for the dilaton black hole. For the second and third term the 
integration over $M$ covers both sides of the domain wall and yield
\be
I_{E_{2}} = {\beta \over 4}\bigg[
\bigg( {Q^2 \over M} - 2 M \bigg) \ln \bigg( {r - {Q^2 \over M} \over r} \bigg)
- 2 {Q^2  \over r} \bigg] {\mid_{r_{+}}}^{r_{st}},
\ee
while
\be
 I_{E_{3}} = {M \over Q^2} \ln \bigg( {r - {Q^2 \over M} \over r} \bigg)   
 {\mid_{r_{+}}}^{r_{st}},
\ee
where $r_{+}$ is the outer horizon of dilaton black hole.\\
In order to get the probability for the pair creation of static charged
dilaton black hole in the presence of the domain wall we divide the 
amplitude for this process by the amplitude for the domain wall creation,
which implies the
following expression $exp( - (I_{st} - I_{0}))$.
In contrast to the process of nucleation of dilaton black holes
on the domain wall, the nucleation in the presence of the domain wall
is characterized by the fact that the mass of black hole is fixed \cite{cal}
and equal to $M = 1/6 \sqrt{3} \pi \sigma$. Therefore it cannot be varied 
independently of the wall's tension. For the extremal dilaton black hole
$P_{dil~ ext} = exp( - (I_{st} - I_{0})) = - 73/ 648 \pi \sigma^2$. If we take
the probabilities for nucleation of Schwarzschild black hole
$P_{Schw} = exp (- 11 / 216 \pi \sigma^2)$ and extremal RN black hole
$P_{RN ~ext} = exp(- 3 /32 \pi \sigma^2)$, one can see that the following takes place:
\be
P_{Schw} > P_{dil~ ext} > P_{RN~ ext}.
\ee
In the considered case the black hole of a certain (large) size can nucleate,
therefore this process will be heavily suppressed. In the nucleation
of black holes on the domain wall the black hole mass is a parameter which can 
be varied independently of $\sigma$, and can be made arbitrarily small. Thus,
the domain walls will prefer to nucleate small black holes on them rather than 
large ones.

\section{Conclusions}
In our work we had studied the problem of dilaton black hole sitting on a domain wall.
Applying the recently deviced C-metric construction \cite{emp1,emp2}
we found the metric for an infinitesimally thin wall intersecting dilaton black hole.
The behavior of the domain wall in the spacetime of the considered black hole
and dilaton C-metric was studied. We derived the {\it thin wall}
approximation useful in our further studies  namely in a gravitational
back reaction problem. Having in mind the behavior of the Abelian-Higgs vortex and extremal
dilaton black hole, we analyzed the domain wall extreme dilaton black hole system
and gave analytic arguments that the extreme dilaton black hole always expelled the domain wall.
Thereby we have extended the phenomenon of the flux expulsion
to the case of dilaton black hole domain wall system. We also considered the 
gravitational back reaction problem concluding that the topological kink solution
smoothed the shell-like singularity of the infinitesimal domain wall.
We studied the nucleation process of dilaton black holes on the domain wall and compared
it to the nucleation of black hole pairs in the presence of a domain wall.
In the last case the black hole  of a certain, large size can be produced.
Therefore it will be heavily suppressed. But in the nucleation of black holes on
the domain wall the black hole mass parameter can be varied independently
of the wall tension and can be made arbitrarily small.
Then, domain walls will rather prefer to nucleate small black holes on them,
than large ones inside them.

\vspace{3cm}
\noindent
{\bf Acknowledgements:}\\
We would like to thank Ruth Gregory
for helpful remarks and discussions on various
occasions. 

\end{document}